\documentclass[
preprint
,showpacs
,showkeys
,nofootinbib
,aps
,amsfonts
,amssymb
,byrevtex
,pdftex 
]{revtex4-1}

\pdfoutput=1
\usepackage{graphicx} 
\usepackage{amsmath, amssymb, bm} 
\usepackage[dvips]{color}

\usepackage{psfrag,epsfig}
\usepackage{natbib}
\usepackage{hyperref}
\usepackage{slashed}

\usepackage[normalem]{ulem}  

\def\){\right)} 
\def\({\left(} 
\def\]{\right]} 
\def\[{\left[}
\def\r{\vec{\bm r}}

\def\nLi{$^7\mathrm{Li}(n, \gamma)^8\mathrm{Li}$}
\def\pBe{$^7\mathrm{Be}(p, \gamma)^8\mathrm{B}$}

\begin{document}

\preprint{INT-PUB-13-029}

\title{Adiabatic projection method for scattering\\ and reactions on the lattice}

\author{%
Michelle Pine$^a$}
\email{mjmantoo@ncsu.edu}

\author{%
Dean Lee$^a$}
\email{dean\_lee@ncsu.edu}

\author{%
Gautam Rupak$^{b}$}
\email{grupak@u.washington.edu}

\affiliation{$^a$ Department of Physics,
North Carolina State University, Raleigh, NC 27695, U.S.A.\\
$^b$ Department of Physics $\&$ Astronomy and 
HPC$^2$ Center for Computational Sciences, 
Mississippi State
University, Mississippi State, MS 39762, U.S.A.}

\begin{abstract}
We demonstrate and test the adiabatic projection method, a general  new framework for calculating scattering and reactions  on the lattice.  The method is based upon calculating a low-energy effective theory for clusters which becomes exact in the limit of large Euclidean projection time.  As a detailed example we calculate the adiabatic two-body Hamiltonian for elastic fermion-dimer scattering  in lattice effective field theory.  Our  calculation corresponds to neutron-deuteron scattering in the spin-quartet channel at leading order in pionless effective field theory.  We show that the spectrum of the adiabatic Hamiltonian reproduces the spectrum of the original Hamiltonian below the inelastic threshold to arbitrary accuracy.  We also show that the calculated $s$-wave phase shift  reproduces the known exact result in the continuum and infinite-volume limits.  When extended to more than one scattering channel, the adiabatic projection method can be used to calculate  inelastic reactions on the lattice in future work.    

\end{abstract}
\maketitle


\section{Introduction}
\label{sec_intro}
\begin{figure}[thb]
\begin{center}
\includegraphics[scale=0.6]{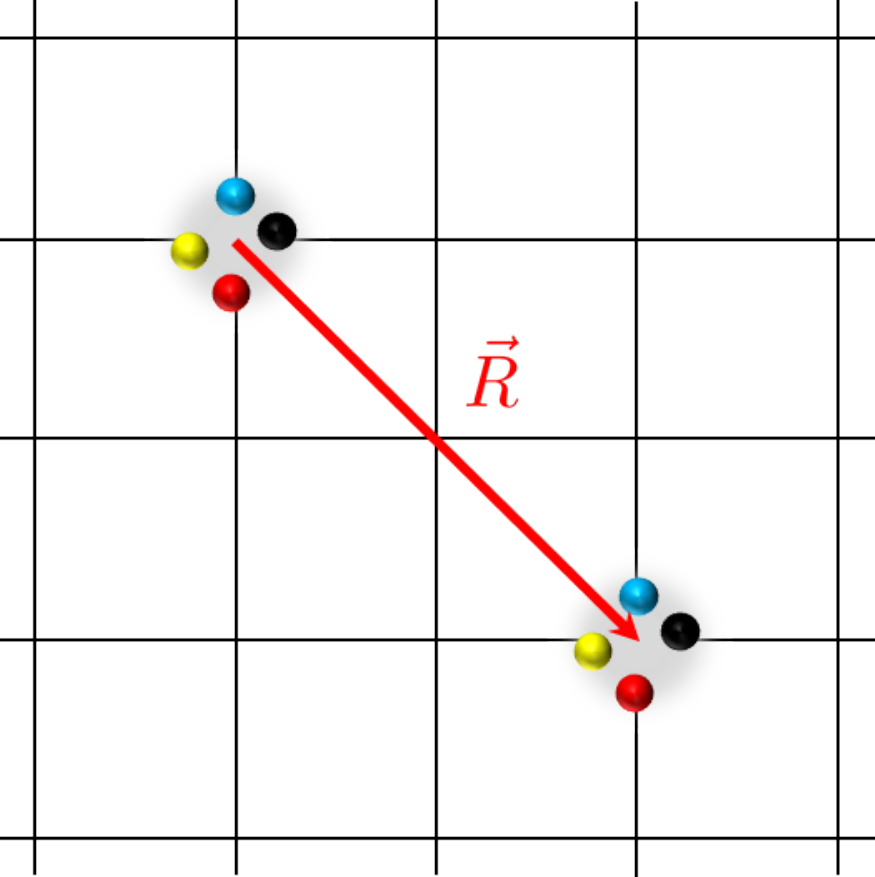}
\caption{Two-body cluster initial state $|\vec{R}\rangle$ separated by displacement vector $\vec{R}$}
\label{fig:clusters}
\end{center}
\end{figure}

There has been recent progress in \emph{ab initio} calculations of nuclear scattering and reactions.  This includes calculations using the
no-core shell model and resonating group method~\cite{Navratil:2011ay,Navratil:2011sa,Navratil:2011zs,Quaglioni:2013kma,Hupin:2013wsa}, fermionic molecular
dynamics~\cite{Neff:2010nm,Neff:2010uk}, 
the coupled-cluster expansion~\cite{Jensen:2010vj,Hagen:2012rq}, and variational and Green's function Monte Carlo~\cite{Nollett:2011qf,Brida:2011yp}.  
For calculations using lattice methods there has been progress in using finite
periodic volumes to analyze coupled-channel scattering~\cite{Liu:2005kr,Lage:2009zv,Bernard:2010fp,Doring:2011ip,Doring:2012eu,Briceno:2012yi} and three-body 
systems~\cite{Polejaeva:2012ut,Briceno:2012rv}. Also the first steps towards calculating
nuclear reactions on the lattice were introduced in Ref.~\cite{Rupak:2013aue} using an adiabatic projection formalism. The
general strategy in the adiabatic projection formalism involves separating the calculation into two parts. The
first part of the method uses Euclidean time projection  to determine an adiabatic Hamiltonian for the participating nuclei.  
This is done by starting with a set of cluster states $|\vec{R}\rangle$ labeled by their separation vector $\vec{R}$, as illustrated in Fig.~\ref{fig:clusters}.  
These states are propagated in Euclidean time to form dressed cluster states,
 \begin{align}
\vert \vec{R}\rangle_\tau =\exp(-H\tau)\vert \vec{R}\rangle .
\end{align}  

By evolving in Euclidean time with the full microscopic Hamiltonian, we are in essence cooling the initial cluster states to the correct physical state dynamically 
with the interaction. Deformations and polarizations of the interacting clusters are incorporated automatically by means of Euclidean time projection. In the limit of large Euclidean time, these dressed cluster states span the  low-energy subspace of   two-body continuum states for our clusters.   We then calculate an adiabatic Hamiltonian matrix defined by the original Hamiltonian restricted to the subspace of dressed cluster states. The construction is still \emph{ab initio} though restricted to the description of cluster configurations.
For inelastic processes we  construct dressed cluster states for each of the possible scattering channels and calculate matrix elements for all operators relevant to the reaction process.  For example, in the case of radiative capture reactions, we calculate the adiabatic Hamiltonian and matrix elements of  one-photon vertex operators.
The second part of the adiabatic projection method entails using the adiabatic Hamiltonian and operator matrix elements for the dressed cluster states to calculate scattering amplitudes.  For elastic phase shifts, we can  apply the finite-volume scaling analysis developed by L\"uscher \cite{Luscher:1986pf,
Luscher:1991ux}.  For inelastic reactions, additional steps in the calculation  are required as follows.  Since we have reduced the problem to a few-body system of nuclear clusters, this part of the calculation can be performed using Green's function methods defined in Minkowskian time. In Ref.~\cite{Rupak:2013aue} this is demonstrated using finite-volume Green's functions for radiative neutron-proton capture, $n + p \rightarrow d + \gamma$, in pionless effective field theory. See Section~\ref{sec_future} for more details. 

In this paper we study in depth the first part of the adiabatic projection method, the construction of dressed cluster states and the adiabatic Hamiltonian.  We use the example of elastic fermion-dimer scattering for attractive two-component fermions in the limit of zero-range interactions.  This corresponds to deuteron-neutron scattering in the spin-quartet channel at leading order in pionless effective field theory.   One of the key requirements of an \emph{ab initio} approach is that all errors are under control and can be systematically reduced.  In our analysis we will find that the spectrum of the adiabatic Hamiltonian matches the spectrum of the original microscopic Hamiltonian below the inelastic threshold to arbitrary accuracy.  Furthermore, we reproduce the $s$-wave phase shift in good agreement with exact results in the continuum and infinite-volume limits  
from the  Skorniakov-Ter-Martirosian (STM) integral equation \cite{Skorniakov:1957,Bedaque:1997qi, Bedaque:1998mb, Braaten:2004a}.
 
The outline of the paper is as follows. In Section~\ref{sec_LEFT} we introduce the underlying interactions for our effective field theory description of two-component fermions and fermion-dimer scattering on the lattice.  In Section \ref{sec_Ha} we apply the adiabatic projection method to this system and calculate the corresponding two-body adiabatic Hamiltonian. In Section \ref{sec_spectrum} we compare the spectrum of the two-body adiabatic Hamiltonian with the spectrum of the original microscopic Hamiltonian.  In Section~\ref{sec_phaseshift} we compute the elastic $s$-wave phase shift for fermion-dimer scattering and compare with exact continuum infinite-volume results. 
 Applications to inelastic processes in future work is discussed in Section~\ref{sec_future}.   
We then conclude in Section~\ref{sec_conclusions} with a summary and outlook.

\section{Two-component fermions on the lattice}
\label{sec_LEFT}

The example we consider in depth is fermion-dimer scattering for two-component fermions.  We call the two components spin-up and spin-down.  The bound dimer state is composed of one spin-up and one spin-down fermion.   The  interactions are chosen to be attractive, and we take the limit where the range of the interactions is negligible.  At leading order in pionless effective field theory,  neutron-deuteron scattering in the spin-quartet channel is completely equivalent to our fermion-dimer scattering system.  In the neutron-deuteron case the two fermion components correspond to isospin,  while all the nucleon intrinsic spins are fully symmetrized into a spin-quartet.

The Hamiltonian for our system can be written as
\begin{align}
H = -\frac{1}{2m}\sum_{i=\uparrow,\downarrow }\int d^{3}r\,a^{\dagger}_{i}(\r)\nabla^{2}a_{i}(\r)+ \int d^{3}r  d^{3} r'\,a^{\dagger}_{\downarrow}(\r)a_{\downarrow}(\r)V(\r-{\r}\,')a^{\dagger}_{\uparrow}(\r\,')a_{\uparrow}(\r\,'),
\end{align}
where we take the zero-range limit $V(\r-{\r}\,')\rightarrow c_0 \delta^{(3)}(\r-{\r}\,')$, with coupling constant $c_0$.  
Here $a_{i}^{\dagger}$ and $a_{i}$ are creation and annihilation operators.   Motivated by the neutron-deuteron system, we take the mass of the fermions to be $m=939$ MeV and tune the strength of $c_0$  to match the binding energy of  the deuteron, $B=2.2246$ MeV.  For a shallow bound dimer such as this, the fermion-dimer scattering problem is known to be strongly coupled even at rather low momenta.  See, for example, Ref.~\cite{Bedaque:1997qi,Bedaque:1998mb} and references therein.   

We will calculate the properties of this system using a lattice Hamiltonian.  While we don't need the full computational machinery of Monte Carlo simulations in this analysis, we should note that the adiabatic projection formalism fits conveniently into the framework of lattice effective field theory.  Lattice effective field theory  is a combination of effective field theory and numerical lattice methods that has been
used to study nuclei in pionless EFT \cite{Borasoy:2005yc} and chiral EFT
\cite{Borasoy:2006qn,Epelbaum:2009zs,Epelbaum:2009pd,Epelbaum:2010xt,Epelbaum:2011md}.  A review of lattice effective field theory calculations can be found in
Ref.~\cite{Lee:2008fa}.

We denote the lattice spacing as $b$.  We write all quantities in lattice units, meaning that we form dimensionless combinations involving the appropriate power of  $b$. Using the simplest possible lattice action with nearest-neighbor hopping terms, we find that the lattice Hamiltonian has the form
\begin{align}
H  =&\frac{1}{2m}\sum_{i=\uparrow,\downarrow}\sum_{l=1}^3 \sum_{\vec{n}}\left[ 2a_{i}^{\dagger}(  \vec
{n})  a_{i}(  \vec{n})-a_{i}%
^{\dagger}(  \vec{n})  a_{i}(  \vec{n}+\hat{l})
+a_{i}^{\dagger}(  \vec{n})  a_{i}(  \vec{n}-\hat{l}) \right] \  
    \nonumber\\
&  +\hat{c}_0\sum_{\vec{n}}a_{\downarrow}^{\dagger}(  \vec{n})
a_{\downarrow}(  \vec{n})  a_{\uparrow}^{\dagger}(  \vec{n})
a_{\uparrow}(  \vec{n})  ,\label{eq:Lattice_Hamiltonian}%
\end{align}
where $\vec{n}$ labels the lattice sites, $\hat{c}_0$ is the lattice-regularized coupling, and $\hat{l}$ is a lattice unit vector in the $l^{\text{th}}$ direction.  We apply cubic periodic boundary conditions, where the physical size of the cube is $L$ times the lattice spacing $b$.

\section{Adiabatic projection method}
\label{sec_Ha}

\begin{figure}[thb]
\begin{center}
\includegraphics[scale=0.6]{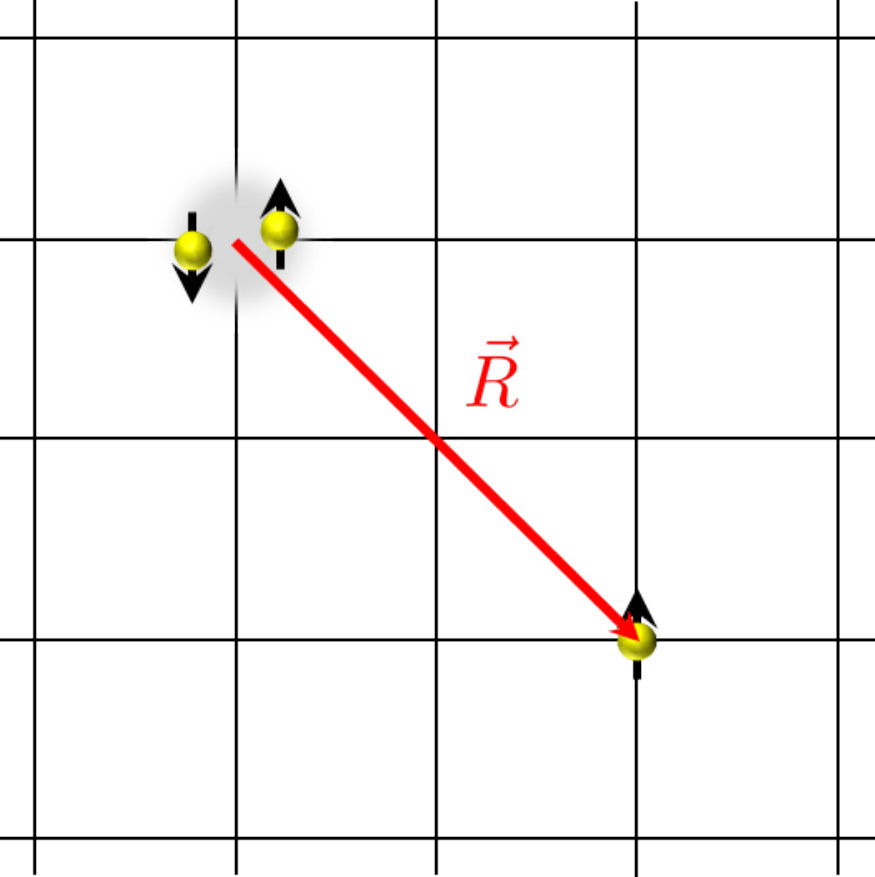}  
\caption{Fermion-dimer cluster initial state $|\vec{R}\rangle$ separated by displacement vector $\vec{R}$}
\label{fig:initial_state}
\end{center}
\end{figure}

The first step of the adiabatic projection method is to set up the initial cluster states.   Without loss of generality we take the fermion-dimer system to consist of two spin-up fermions and one spin-down fermion.  We will work in the center-of-mass frame and measure particle locations relative to the spin-down fermion.  In our coordinate convention the spin-down fermion is anchored at the origin, $\vec{0}$, while the two spin-up fermion locations are unconstrained except for Fermi statistics.  We choose  our cluster initial states to have the form 
 \begin{align}
|\vec{R} \rangle= a_\uparrow^\dagger(\vec{R} ) a_\uparrow^\dagger(\vec{0}) a_\downarrow^\dagger(\vec{0})|0\rangle
\label{R}
\end{align}
for any lattice separation vector $\vec{R} \ne \vec{0}$.  This is illustrated in Fig.~\ref{fig:initial_state}.  
In the actual code we use a Slater determinant to construct the fermionic state. 
On our cubic periodic lattice, there are $L^3-1$  possible values for $\vec{R}$. 
We now evolve the initial states $|\vec{R}\rangle$ in Euclidean time $\tau$ with the microscopic Hamiltonian to produce the dressed cluster states,
\begin{align}
|\vec{R}\rangle_\tau=\exp(-H\tau)|\vec{R}\rangle.
\end{align}
We have chosen a  simple form for the initial cluster states in Eq.~(\ref{R}) to demonstrate the general properties of the adiabatic projection method as simply as possible.  We can accelerate the convergence of the method by choosing an initial cluster state that better reproduces the dimer wavefunction. The Euclidean time evolution is done by exact matrix multiplication using the Trotter approximation
\begin{align}
\exp(-H\tau)\approx [1-\frac{\tau}{L_\tau} H]^{L_\tau},
\end{align}
for large number of time steps $L_\tau$. 

The initial state $|\vec{R}\rangle$ plays a role analogous to  an interpolating field.  We start with a configuration which roughly approximates the desired continuum state.  The Euclidean time projection  then systematically improves the approximation while accounting for all possible deformations and polarizations due to the interacting bodies.  In the limit of large projection time $\tau$, the set of  dressed cluster states $|\vec{R}\rangle_{\tau}$ will span the low-energy spectrum of the original Hamiltonian $H$.

The technique of generating cluster scattering states using Euclidean 
time projection is motivated by recent studies of alpha-particle clusters in the carbon-12 nucleus~\cite{Epelbaum:2011md,Epelbaum:2012qn}.  In those investigations, two different characteristic time scales are apparent from the projection Monte Carlo simulations.  The first is a fast time scale associated with the formation of alpha clusters.  Starting from any initial state of carbon-12, individual clusters emerge quickly as a function of projection time $\tau$.  However the overall structure of the alpha clusters relative to each other develops only much later in projection time $\tau$.   The underlying physics is related to the original motivation of Wheeler when he first introduced the resonating group method  to describe the structure of compound nuclei \cite{Wheeler:1937zz}.

The same separation of time scales can be seen in the Euclidean time projection of continuum states.  The formation time for individual clusters is fast while the physics of inter-cluster interactions develops more slowly. The adiabatic projection formalism uses this separation of time scales to represent the low-energy continuum states efficiently as superpositions of dressed cluster states.

The dressed cluster states  $|\vec{R}\rangle_\tau$ are in general not orthogonal.  Therefore it is convenient to define the dual vector $_{\tau}(\vec{R}\vert$  as a
linear functional,
\begin{align}
_{\tau}(\vec{R}\vert v\rangle\ \equiv\ \sum_{\vec{R}'}\left[ N_{\tau}^{-1} \right]_{\vec{R},\vec{R}'} {}_{\tau}\langle\vec{R}'\vert v\rangle,
\label{eq:dual}
\end{align}
where $N_{\tau}^{-1}$ is the inverse of the norm matrix $N_{\tau}$ with components given by the inner product
\begin{align}
\left[N_{\tau}\right]_{\vec{R},\vec{R}'} =\ _{\tau}\langle\vec{R}\vert\vec{R}'\rangle_{\tau}.
\end{align}
The dual vector $_{\tau}(\vec{R}\vert$ will annihilate any vector which is orthogonal to all dressed cluster states: 
\begin{align}
 {}_{\tau}\langle\vec{R} \vert v\rangle=0 \text{ for all }  \vec{R} \,\Rightarrow \,_{\tau}(\vec{R}\vert v\rangle=0  \text{ for all }  \vec{R}.
\end{align}
It also serves as a dual basis within the linear subspace of dressed cluster states, 
\begin{align}
_{\tau}(\vec{R}\vert \vec{R}'\rangle_\tau = \delta_{\vec{R},\vec{R}'} .
\end{align}

Let $H^{a}_{\tau}$ be the matrix representation of the Hamiltonian operator $H$ projected onto the set of dressed cluster states,
\begin{align}
\left[ H^a_{\tau} \right]_{\vec{R},\vec{R}'} = \ _{\tau}(\vec{R}\vert H \vert \vec{R}'\rangle_{\tau}.
\end{align}
We will call $H^{a}_{\tau}$ the two-body adiabatic Hamiltonian, an effective two-body matrix Hamiltonian describing the fermion-dimer system.  From Eq.~(\ref{eq:dual}) we can write the adiabatic Hamiltonian as 
\begin{align}
\left[ H^a_{\tau} \right]_{\vec{R},\vec{R}'} = \sum_{\vec{R}''}\left[ N_{\tau}^{-1} \right]_{\vec{R},\vec{R}''} {}_{\tau}\langle\vec{R}''\vert H \vert\vec{R}' \rangle_{\tau}.
\end{align}
If we apply a similarity transform involving the inverse square root of the norm matrix, $N^{-1/2}_{\tau}$, then in the new basis the adiabatic Hamiltonian is Hermitian,
\begin{align}
\left[ {H^a_{\tau}}' \right]_{\vec{R},\vec{R}'} = \ \sum_{\vec{R}'',\vec{R}'''}
\left[ N_{\tau}^{-1/2} \right]_{\vec{R},\vec{R}''} {}_{\tau}\langle\vec{R}''\vert H \vert\vec{R}''' \rangle_{\tau}\left[ N_{\tau}^{-1/2} \right]_{\vec{R}''',\vec{R}'}.
\end{align}
The structure of this Hermitian adiabatic Hamiltonian is similar to the Hamiltonian matrix  used in recent calculations of the no-core shell model together with the resonating
 group method~\cite{Navratil:2011ay,Navratil:2011sa,Navratil:2011zs,Quaglioni:2013kma,Hupin:2013wsa}.

\section{Results for the Finite-Volume Spectrum}
\label{sec_spectrum} 

Let the low-energy spectrum of the microscopic $H$ be denoted 
\begin{align}
E_0\leqslant E_1\leqslant E_2\cdots.
\end{align}  
Let $N_R$ be the number of initial cluster states that we use in our calculation. It is not necessary to use every possible cluster state on the lattice, and we will discuss the choice of $N_R$ a bit later in our discussion.  Suppose now we construct an adiabatic Hamiltonian $H^a_{\tau}$ defined in  the subspace that is spanned by $N_R$ cluster separation states $\vert \vec{R}\rangle$. In the asymptotic limit $\tau \rightarrow \infty$, it is straightforward to prove that the spectrum of $H^a_{\tau}$ will match the low-energy spectrum 
 $E_0,\cdots E_{N_R-1}$ with an error
\begin{align}
\Delta E_j \sim O\{\exp{\left[- 2\left( E_{N_R} - E_j \right)\tau \right]\}}.
\label{eq:error}
\end{align} 
In most practical applications, however, we cannot go to extremely large values for $\tau$.  
Therefore we actually see a more complicated dependence on $\tau$ associated with higher-body 
continuum states.  

At finite volume and above the threshold for three-body states, the eigenstates of $H$ will in general be a mixture of two-body and three-body states.  However at large volumes we can still classify which energy eigenstates are predominantly two-body or predominantly three-body.  Our initial two-body cluster states $\vert \vec{R} \rangle$ will have only a very small overlap with the three-body continuum states.  In the adiabatic projection method we would need to include initial states that have better overlap with three-body states in order to reproduce the three-body continuum spectrum of $H$.

Consider for the moment the idealized case where our initial states $\vert \vec{R} \rangle$ are completely orthogonal to all three-body and higher-body states.  Let 
 $ E^{(2)}_0,\cdots E^{(2)}_{N_R-1}$ be the energy-ordered spectrum of $H$ including up to at most two-body states.  We then reproduce the energy level $E^{(2)}_j$ using the adiabatic projection method with error
 \begin{align}
\Delta E^{(2)}_j \sim O\left\{\exp{\left[- 2\left( E^{(2)}_{N_R} - E^{(2)}_j \right)\tau \right]}\right\}.
\end{align}

Now let us return to the actual situation where there is some small but nonzero overlap between our initial states $|\vec{R} \rangle$ and higher-body states.  These higher-body states introduce a small additional error to the calculation of the two-body energy levels, 
\begin{align}\label{eq:three-body}
\Delta E^{(2)}_j \sim O\left\{\exp{\left[- 2\left( E^{(2)}_{N_R} - E^{(2)}_j \right)\tau \right]}\right\}+\sum_{E _{}}\sigma^{(\geqslant3)}_ j(E) \exp{\left[- 2\left( E - E^{(2)}_j \right)\tau \right]}.
\end{align}
Here $\sigma^{(\geqslant3)}_ j(E)$ denotes an energy-dependent spectral function which characterizes the small overlap between the higher-body states and our two-body cluster states 
$\vert \vec{R} \rangle$.

We should mention that it is neither necessary nor advantageous to include all possible lattice separation vectors $\vec{R}$ in the set of initial cluster states $\vert \vec{R} \rangle$.  It is sufficient and often more efficient to keep a smaller set of vectors. This technique is useful as it significantly reduces the numerical task of computing  $|\vec{R}\rangle_\tau$ and avoids numerical stability problems produced by large ill-conditioned norm matrices $N_{\tau}$.  Skipping states $\vert \vec{R} \rangle$ will reduce $N_R$ and this in turn will decrease the energy level $E^{(2)}_{N_R}$.  The effect on the convergence of the method as a function of projection time $\tau$ can be understood from the error estimate in Eq.~(\ref{eq:three-body}).  Roughly speaking it is most  efficient to choose a value for $N_R$ which is somewhat larger than the number of scattering states we would like to compute.  The strategy then is to choose $\tau$ sufficiently large so that the desired accuracy goal is achieved.

\begin{figure}[thb]
\begin{center}
\includegraphics[width=0.70\textwidth]{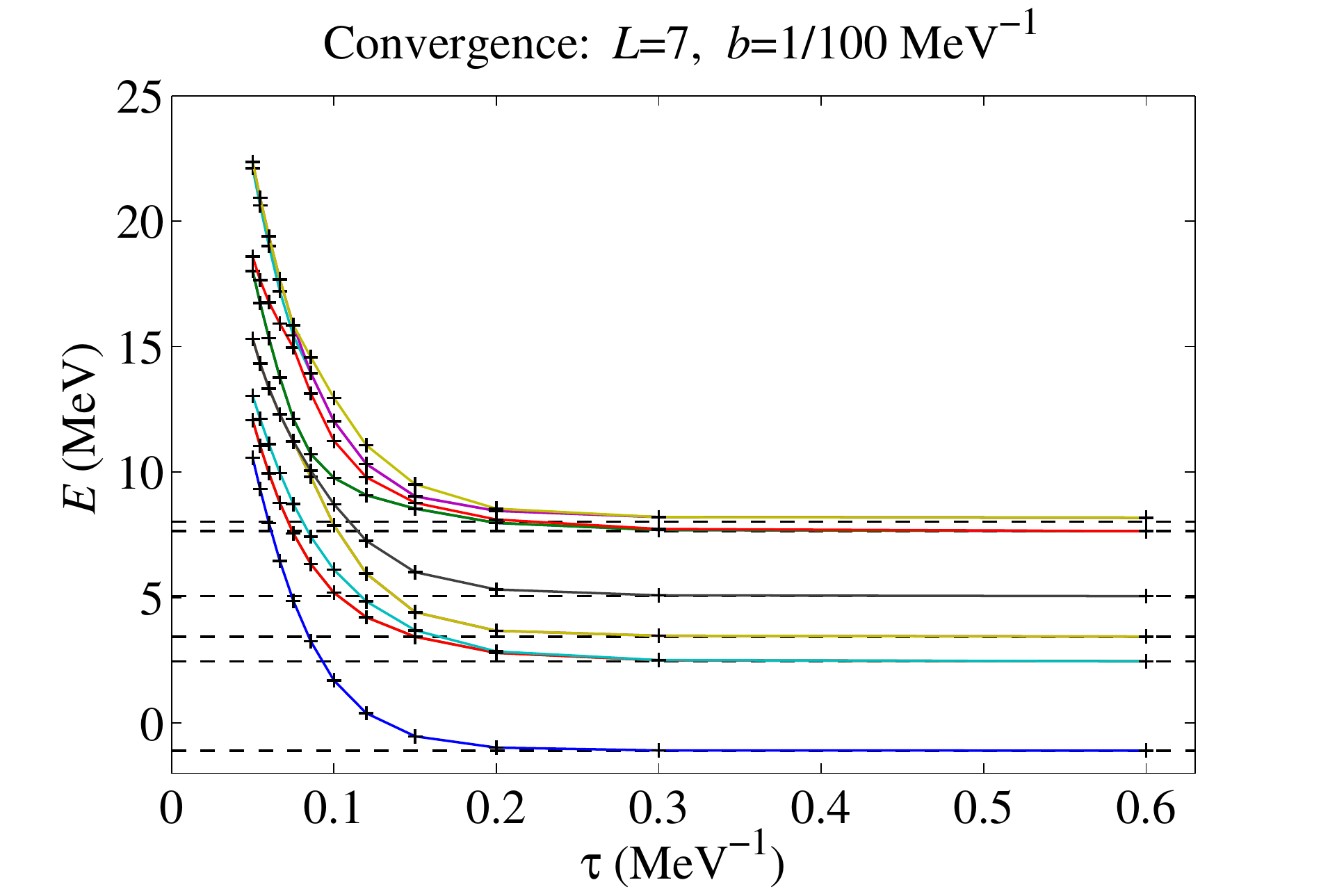} 
\caption{\protect Energy levels  versus projection time $\tau$  for the adiabatic Hamiltonian compared with energy levels for the microscopic Hamiltonian.}
\label{fig:convergence}
\end{center}
\end{figure}

In Fig.~\ref{fig:convergence} we  plot the low-lying energy levels of $H^{a}_{\tau}$ as a function of Euclidean time $\tau$.  We show results for lattice spacing $b=1/100$ MeV$^{-1}$ and cubic periodic box size $L=7$.  This corresponds with a physical length of $Lb = 13.8~\text{fm}$.  For these calculations we use initial cluster states $\vert \vec{R} \rangle$ which are separated by at least two lattice sites  in each direction.  The energy levels of the microscopic Hamiltonian $H$ are indicated by the horizontal lines. As the system evolves in Euclidean time, the lowest ten energy levels of the adiabatic Hamiltonian fall onto the lowest ten energy levels of $H$ with exponential convergence.  The degeneracies of these levels are not shown in Fig.~\ref{fig:convergence}, but these first ten eigenvalues correspond to five  energy levels.  For this  chosen periodic box volume, the spectrum of predominantly three-body continuum states starts at $E = 8.0~\text{MeV}$, the topmost horizontal line shown in Fig.~\ref{fig:convergence}.  We identify the different continuum states by measuring spatial correlations among the three fermions. In particular, the fermion-dimer states are easily distinguished due to their significant probability for a spin-up and spin-down fermion occupying the same lattice site.

As expected, the three-body continuum states are not accurately reproduced for the values of $\tau$ shown in Fig.~\ref{fig:convergence}. However, the two-body continuum states below the three-body threshold are produced with rapid exponential convergence.  The exponential dependence of the errors are consistent with spectral functions as defined in Eq.~(\ref{eq:three-body}) that are peaked at energies slightly above the three-body threshold at $8.0~\text{MeV}$.
As the box size increases, the number of two-body continuum states below the three-body threshold increases.  Therefore we will reproduce more two-body continuum states using the two-body adiabatic Hamiltonian at larger volumes.

\section{Adiabatic Projection Calculation of s-wave Phase Shift}
\label{sec_phaseshift}

We now use adiabatic projection to calculate the fermion-dimer elastic $s$-wave phase shift $\delta_0$.  We will extract the phase shift at finite volume using L\"uscher's finite-volume method \cite{Luscher:1986pf,Luscher:1991ux}.  We consider three lattice spacings $b=1/100$  MeV$^{-1}$ , 1/150 MeV$^{-1}$, 1/200 MeV$^{-1}$ and a range of lattice box sizes $L\leq 16$.  This corresponds to maximum physical box sizes of $32$ fm, 21 fm  and  $16$ fm, respectively.  

We will work in the center-of-mass frame of the fermion-dimer system.  The phase shift $\delta_0$ is calculated from the low-energy spectrum in the cubic periodic box using the relation
\begin{align}\label{eq:Luscher}
p\cot\delta_0(p) =\frac{1}{\pi L }S(\eta),\quad \eta = \left( \frac{pL}{2\pi} \right)^2, 
\end{align}
where  $p$ is the relative momentum between the two scattering bodies  as deduced from the finite-volume fermion-dimer energy $E^\mathrm{fd}$, and $S$ is the three-dimensional zeta function, 
\begin{align}
S(\eta) = \lim_{\Lambda \to \infty}\left[\sum_{\vec{n}}\frac{\theta(\Lambda^{2}-\vec{n}^{2})}{\vec{n}^{2}-\eta}-4\pi\Lambda\right].
\end{align}
All quantities are written in terms of lattice units. The physics of the scattering process is encoded in the discrete values of the momentum $p$ set by the energy levels in the periodic cube.  

The fermion-dimer energy in the periodic box of size $L b$ for $s$-wave scattering states can be written as~\cite{Bour:2011ef,Bour:2012hn,Rokash:2013xda}
\begin{align}\label{eq:Efd_L}
E^\mathrm{fd}(p,L)= E^\mathrm{fd}(p,\infty)+\tau^{\text{d}}(\eta)\Delta E_{\vec{0}}^{\text{d}}(L).
\end{align}
The lattice momentum $p$  in Eq.~(\ref{eq:Luscher}) is self-consistently derived from the above equation using the various expressions as described below. 

The first term on the right hand side of Eq.~(\ref{eq:Efd_L}) is the infinite-volume fermion-dimer energy given by
\begin{align}
E^\mathrm{fd}(p,\infty)=\frac{p^2}{2m_d} +\frac{p^2}{2m} -B(\infty),
\end{align}
with  $m_{\text{d}}$ the dimer effective mass, and $B(\infty)\equiv B=2.2246$ MeV the infinite-volume dimer binding energy.  In the non-relativistic continuum limit, the dimer mass $m_{\text{d}}$ equals $2m$. However, in order to reduce systematic errors in our lattice calculation, we take into account the renormalization of the dimer effective mass
$m_{\text{d}}$ at nonzero lattice spacing $b$.   We determine the dimer effective mass by numerically calculating the dispersion relation of the dimer on the lattice using a very large volume, $L=60$, in order to eliminate any finite-volume effects. 

The second term  $\tau^{\text{d}}(\eta)\Delta E_{\vec{0}}^{\text{d}}(L)$ in Eq.~(\ref{eq:Efd_L}) encapsulates the finite-volume corrections to the fermion-dimer energy due to the dimer wavefunction wrapping around the periodic boundary~\cite{Luscher:1985dn,Konig:2011nz,Konig:2011ti}, where   $\Delta E_{\vec{0}}^{\text{d}}(L)=B(\infty)-B(L)$ is the finite-volume energy shift for the bound dimer state in the two-body center-of-mass  frame.  The topological factor   $\tau^{\text{d}}(\eta)$ for the $s$-wave dimer wavefunction wrapping around the periodic boundary once is given by~\cite{Bour:2012hn}
\begin{align}\label{eq:taud}
\tau^{\text{d}}(\eta) = 
\frac{1}{\mathcal{N}}\sum_{\vec{k}}\frac{\tau(\vec{k},1/2)}{(\vec{k}^2-\eta^{2})^{2}} ,\quad \mathfrak{\mathcal{N}} = \sum_{\vec{k}}\frac{1}{(\vec{k}^{2}-\eta^{2})^2},
\end{align}
where 
\begin{align}
\tau(\vec{k},1/2)=\frac{1}{3}\sum_{i=1}^3\cos(k_iL/2),
\end{align}
accounts for the finite-volume energy shift  of a dimer moving with center-of-mass momentum $\vec{k}$ in the fermion-dimer system
 \begin{align}
\Delta E^{\text{d}}_{\vec{k}}(L)=\tau(\vec{k},1/2)\Delta E^{\text{d}}_{\vec{0}}(L). 
\end{align}
The expression for the topological factor $\tau^{\text{d}}(\eta)$ in Eq.~(\ref{eq:taud}) ignores higher-order volume corrections associated with the dimer wavefunction winding around the boundary more than once and finite-volume corrections to the fermion-dimer interactions. 

From low energy spectrum in the lattice calculation, we directly determine $E^{fd}(L)$ and $\Delta E_{\vec{0}}^{\text{d}}(L)= B-B(L)$.  Then the  lattice momentum $p$ (and $\eta = \left( \frac{pL}{2\pi} \right)^2$) is the one that satisfied Eq.~(\ref{eq:Efd_L}) with these energies. We solve for $p$ iteratively by setting $\tau^{\text{d}}(\eta)=1$ at the start, and then recursively solving for $\tau^{\text{d}}(\eta)$ and $p$ until they  convergence.  Note $\tau^{\text{d}}(\eta)=1$ gives $\Delta E^{\text{d}}_{\vec{k}}(L)=\Delta E^{\text{d}}_{\vec{0}}(L)$ that corresponds to
\begin{align}
E^\mathrm{fd}(p,L)=\frac{p^2}{2m_d} +\frac{p^2}{2m} -B(L),
\end{align} 
a naive but reasonable initial guess for the lattice momentum $p$ given $E^\mathrm{fd}(L)$ and $B(L)$ from the lattice calculation.   In our calculations, we find convergence within approximately 10-15 iterations~\cite{Bour:2012hn}. Having determined $p$, the phase shift is calculated from Eq.~(\ref{eq:Luscher}).

We now compare the lattice calculation of the fermion-dimer $s$-wave phase shift to the exact result in the continuum and infinite volume limits obtained from  the STM integral equation.  For the $s$-wave  half-off-shell fermion-dimer scattering $T$-matrix,   we find~\cite{Bedaque:1999vb, Gabbiani:1999yv,Rupak:2001ci} 
\begin{align}
T(k,p)=&-\frac{4\pi\gamma}{m k p} \ln\[\frac{k^2+p^2+k p -m E}{k^2+p^2-k p -m E}\]\nonumber\\
&-\frac{1}{\pi}\int_0^\infty dq \(\frac{q}{p}\)
\frac{T(k,q)}{-\gamma+\sqrt{3 q^2/4- m E-i 0^+}}  \ln\[\frac{q^2+p^2+q p -m E}{q^2+p^2-q p -m E}\],
\end{align}
with dimer binding momentum $\gamma=\sqrt{m B}$ and total energy
$E=3 p^2/(4 m) -B$. 
We determine the phase shift from the on-shell $T$-matrix,
\begin{align}
iT(p,p)= \frac{3\pi}{m}\frac{i}{p\cot\delta_0-ip} .
\end{align}

\begin{figure}[thb]
\begin{center}
\includegraphics[width=0.65\textwidth]{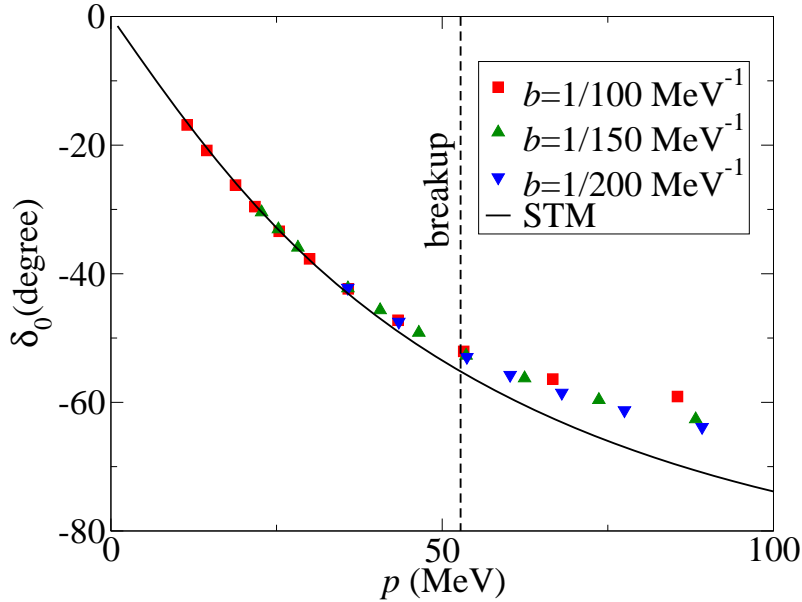}
\caption{\protect Plot of the $s$-wave elastic phase shift.  Solid curve is the STM result.  The squares show  lattice data at  $b=1/100$ MeV$^{-1}$, upright triangles show lattice data at $b=1/150$ MeV$^{-1}$, and upside-down triangles show lattice data at $b=1/200$ MeV$^{-1}$. The dimer breakup momentum at 52.7 MeV is indicated by the vertical dashed line. 
}\label{fig:phaseshift}\end{center}\end{figure}

In Fig.~\ref{fig:phaseshift} we compare the lattice results and STM equation results for the $s$-wave phase shift.  We show data for three different lattice spacings $b=1/100$  MeV$^{-1}$, $b=1/150$ MeV$^{-1}$ and $b=1/200$ MeV$^{-1}$. For each of these adiabatic projection calculations we use about thirty initial cluster states, $N_R \sim 30$.  At low momentum the lattice results are expected to be accurate, and this is evident in the plotted results.  
There is only a small deviation starting near the dimer breakup momentum 52.7 MeV.   The three-body breakup amplitude can also  be calculated using the adiabatic projection formalism.  However this requires the inclusion of low-lying three-body states and is beyond the scope of our analysis here.  Investigations on this topic are planned in future work. The fermion-dimer breakup amplitude happens to be numerically small   at low momenta.  Consequently we see reasonable agreement between the lattice and STM results for the elastic phase shift even above the breakup momentum. 

\section{Future Directions and Extensions}
\label{sec_future}
In order to calculate two-body scattering processes with charge or mass transfer, we can generalize the formalism to include an additional scattering channel.  This is done by using two sets  of initial cluster states $\vert\vec{R}^{1}\rangle$ and $\vert\vec{R}^{2}\rangle$.  We then consider  the dressed cluster states for each channel and their mixing  to compute the multi-channel norm matrix, 
\begin{align}
\left[ N^{}_{\tau} \right]_{\vec{R}^{i},\vec{R}'^{j}}=\,_{\tau}\langle\vec{R}^i\vert\vec{R}'^j\rangle_{\tau}
\end{align}
and multi-channel adiabatic Hamiltonian,
\begin{align}
\left[ H^{a}_{\tau} \right]_{\vec{R}^{i},\vec{R}'^{j}} = \sum_{\vec{R}''^{k}}\left[ N_{\tau}^{-1} \right]_{\vec{R}^{i},\vec{R}''^{k}} {}_{\tau}\langle\vec{R}''^{k}\vert H \vert\vec{R}'^{j} \rangle_{\tau}.
\end{align}
These inelastic scattering processes will be investigated in future studies.

Another interesting and important application of the adiabatic projection method is radiative capture.   Radiative capture reactions have great relevance to understanding hydrogen and helium burning in stars.  The determination of  astrophysical $S$-factors and asymptotic normalization coefficients are important areas where more theoretical input is needed.  For example, model-independent analyses of \nLi ~and \pBe ~at low energies indicate that the strong nuclear interaction component of this process~is sensitive to the elastic $n$-$^7$Li and 
$p$-$^7$Be scattering parameters at leading order and these are not well constrained experimentally~\cite{Rupak:2011nk,Fernando:2011ts}.

To calculate radiative capture reactions in the adiabatic projection formalism, we need to add a one-body cluster state $\vert X\rangle$ to the set of initial cluster states.  This state will correspond to the outgoing nucleus after capture.  We again compute the corresponding multi-channel norm matrix and multi-channel adiabatic Hamiltonian.  In this case, though, we also need to compute one-photon transition matrix elements between the dressed cluster states
\begin{align}
_{\tau}\langle X \vert O_{\gamma}\vert \vec{R} \rangle_{\tau}.
\end{align}
After computing all of the quantities involving the dressed cluster states, we have now reduced the problem to radiative  capture involving only two incoming bodies.  Hence the capture amplitude can be calculated in the same manner as calculating radiative neutron-proton capture, $n + p \rightarrow d + \gamma$.  This is demonstrated in detail on the lattice in Ref.~\cite{Rupak:2013aue} using infrared-regulated Green's function methods.

\section{Summary and Outlook}
\label{sec_conclusions}

In this paper we have demonstrated and tested  the adiabatic projection method, a general  framework for calculating scattering and reactions  on the lattice.  The adiabatic projection method is based upon computing a low-energy effective theory for clusters.  In our analysis we calculated the adiabatic two-body Hamiltonian for elastic fermion-dimer scattering with zero-range attractive two-component fermions. This system corresponds to neutron-deuteron  scattering in the quartet channel at leading order in pionless effective field theory. Future work should include higher order corrections in the effective field theory. 

We found that the spectrum of the two-body adiabatic Hamiltonian matches the low-energy spectrum of the fermion-dimer
system below the three-body continuum threshold.  In the limit of large projection time, the adiabatic Hamiltonian  estimates for the energy levels become exact, and the errors are exponentially small in the projection time $\tau$.  Using L\"uscher's finite-volume method, we found good agreement between lattice results and continuum STM equation results for the $s$-wave phase shift up to the dimer breakup threshold.

While we did not employ Monte Carlo methods in our calculation here, the adiabatic projection method uses Euclidean time projection and is therefore compatible with large-scale projection Monte Carlo codes being used in lattice effective field theory calculations.  In particular, the Euclidean time projection can be performed using the same auxiliary-field projection Monte Carlo method used in previous calculations.  These have the advantage of very favorable scaling with particle number when 
the sign problem is not severe.  The computational effort scales with nucleon number roughly as $A^2$  for $A$ in the range of about twenty nucleons.

The scattering calculations, however, require more work than simple bound state calculations.  If we use $N_R$ initial cluster states, then the calculation is at least a factor of $N_R$ times more in computational effort.  In addition to this, we would also need to explore volumes larger than the box sizes used in bound state calculations. However the overall difficulty of each individual Monte Carlo calculation is not significantly greater than that for bound state calculations.  We are currently working on the application of Green's function methods to calculate scattering amplitudes for elastic and inelastic processes by computing $G$-matrix elements.  This should eliminate problems associated with L\"uscher's finite-volume method and the need for high-accuracy calculations of finite-volume energy levels. Benchmark calculations of the adiabatic projection method using Monte Carlo simulations will be presented in several publications in the near future.

\begin{acknowledgments}
The authors thank U.-G. Mei{\ss}ner for valuable comments on the manuscript.  
Computing support was provided by the HPCC at MSU. Part of this work was completed at the Institute for Nuclear Theory, Seattle.  The authors thank E. Epelbaum and H. Krebs for kind hospitality at Ruhr-Universit\"{a}t, Bochum.  Partial support provided by the U.S. Department of Energy grant DE-FG02-03ER41260 (D.L. and M.P.), U.S. Department of Education GAANN Fellowship (M.P.),  and  the U.S. National Science Foundation  grant No. PHY-0969378 (G.R.)
\end{acknowledgments}

\bibliographystyle{apsrev4-1}
\bibliography{Reference}
\end{document}